\documentclass[showpacs,preprintnumbers,amsmath,amssymb]{revtex4}
\usepackage{epsfig,float,wrapfig,amsmath,graphicx}
\newcommand{\be}{\begin{equation}}
\newcommand{\ee}{\end{equation}}
\newcommand{\beq}{\begin{eqnarray}}
\newcommand{\eeq}{\end{eqnarray}}

\begin{document}

\title{Angular distribution and azimuthal asymmetry for 
pentaquark production in proton-proton collisions}

\author{H.W. Barz}
\affiliation{Institut f\"ur Kern- und Hadronenphysik, 
Forschungszentrum Rossendorf, Pf 510119, D-01314 Dresden, Germany}
\author{M. Z\'et\'enyi}
\affiliation{KFKI Research Institute for Particle and Nuclear Physics, POB 49,
H-1525 Budapest, Hungary}

\date{\today}

\begin{abstract}
Angular distributions for production of the $\Theta^+$ pentaquark
are calculated for the collisions of polarized protons with
polarized target protons. We compare calculations based on 
different assumptions concerning spin and parity ($J=1/2^\pm,3/2^\pm$)
of the $\Theta^+$ state.
For a wide class of interactions the spin correlation parameters 
describing the asymmetric angular distributions are calculated
up to 250 MeV above production threshold. The deviations from the
near threshold behavior are investigated.
\end{abstract}


\maketitle


\section{Introduction}

Recently a number of experiments \cite{LEPS,DIANA,CLAS1,CLAS2,SAPHIR,
HERMES,ZEUS}
have confirmed the existence of a narrow pentaquark state $\Theta^+$ with a 
mass of about 1.53 GeV.
It decays into the $pK^0$ or the $nK^+$ channel with a width
of less than about 15 MeV. Its positive strangeness $S=+1$
and the small width give good reason to identify this resonance 
with the pentaquark state
predicted in ref.~\cite{Diakonov} within the chiral soliton
model. According to this approach it belongs to a
$J^\pi = 1/2^+$  antidecuplet as an isospin singlet with the
five quark configuration $uudd\bar{s}$. 
However, the spin-parity assignment is experimentally still not verified.
First measurements are done to observe this resonance in proton-proton
collision $pp \to p \Sigma^+ K^0$ \cite{COSY} at near threshold energy
where the invariant masses of the $p K^0$ pair indicate the existence of 
the $\Theta^+$ resonance.

Several investigations are made to find observables to determine
spin and parity.  In photo-production processes 
the cross sections were studied using $K$ \cite{Nam2,LiuKo}and 
$K$ and $K^*$ \cite{nakayama,close2} exchange Lagrangians.
For nucleon-nucleon reactions the cross section \cite{LiuKo} and 
the azimuthal angular distribution \cite{Hanhart,Nam1} 
were analyzed. 
Especially in the near threshold region the azimuthal angular
distribution is very sensitive to 
distinguish between different parity assignments of the 
$\Theta^+$ particle. The great advantage of using the threshold
region lies in the fact that to a large extent model independent
predictions can be made \cite{Hanhart,thomas,uzikov,rekalo}.

In this work we are going to extend the study of proton-proton
reactions to larger energies using a model dependent interaction.
We employ a combination of $K$ and  $K^*$
exchange, where the parameters are constraint by assuming a production
cross section as was found in the experiment \cite{COSY}.
The azimuthal angular distribution is parametrized 
by correlation coefficients, which we calculate assuming four 
different spin-parity states $1/2^\pm$, $3/2^\pm$ of the 
$\Theta^+$ resonance. It is the aim to investigate how the
threshold behavior of the azimuthal angular distribution changes
if the energy is increased above threshold.

\section {The model}

The simplest mechanism than can be used to  describe the reaction
$pp\to \Sigma^+ \Theta^+$ is the exchange of a pseudoscalar kaon
given by the interaction Lagrangian

\beq  \label{theta1}
{\cal{L}}_{KN\Theta}& =&\;i g_{KN\Theta} \overline{\Theta^+}
                      \gamma^5 (pK^0 + nK^+) \, + h.c. \;, \\
{\cal{L}}_{KN\Sigma}& =&\; i \sqrt{2} \, 
                      g_{KN\Sigma} \overline{\Sigma^+}
		      \gamma^5 (p \overline{K^0}  +n \overline{K^+})
		      \, + h.c.
      \label{sigma}
\eeq
for the $1/2^+$ state of the $\Theta^+$ particle.
The symbols $\Theta^+$, $\Sigma^+$, $p$, $n$ stand for the 
spinors of the participating Fermions and $K^\pm$ for the bosonic 
wave functions. The factor $\sqrt{2}$ in Eq.~(\ref{sigma}) comes 
from the isospin factor in the
standard representation \cite{LiKo}.
The coupling constant $g_{KN\Theta}$ is related to the decay 
width of the $\Theta^+$ into to the $K^+$ and $K^0$ channels via
\be   \label{gamma1}
    \Gamma_\Theta = g_{KN\Theta}^2 \frac{p_N (p^0_N -m_N)}
                    {2\pi m_\Theta},
\ee
where $p_N$ denotes the momentum of the emitted proton with 
$p^0_N$   being the
energy as the zeroth component.
Assuming a width of 10 MeV and a mass of $\Theta^+$ of $m_{\Theta}=1.53$ GeV
we obtain $g_{KN\Theta}$ = 3.27.
(Another possible approach for the determination of $g_{KN\Theta}$ uses SU(3)
relations of the pentaquark multiplets \cite{OhKimLee}. Assuming that the
N(1710) baryon resonance is an ideal mixture of antidecuplet and octet
pentaquark states, they obtain $g_{KN\Theta}$ = 3.0 \cite{LiuKoKub}.)
The coupling constant $g_{KN\Sigma}$ was estimated within the framework
of SU(3) \cite{LiuKo,LiKo} to be $g_{KN\Sigma}$=-3.78. (The actual value
depends somewhat on  the data the SU(3) parameters 
are adjusted to.) Furthermore we use 
a monopol formfactor 
\be 
F(q^2) \,=\, \frac{\Lambda^2 - m_K^2} { \Lambda^2 - q^2}
\ee
with 
 $ q$ being the square of the transferred four-momentum.

In the following we also consider the possibilities that spin and parity
of the $\Theta^+$ could
take the 
values $J^\pi$ = $1/2^-$, $3/2^+$, or $3/2^-$. The corresponding 
Lagrangians in their simplest form read
\beq  \label{theta2}
{\cal{L}'}_{KN\Theta}\,& =&\,  
            g'_{KN\Theta} \overline{\Theta^+} (pK^0 + nK^+)  
		      + h.c.\,, \\
 \label{theta3}
{\cal{L}''}_{KN\Theta}\, &=&\,   
            \frac{f_{KN\Theta}}{m_K} \overline{\Theta^+_\mu} 
	    (p\partial^\mu K^0 + n\partial^\mu K^+)  + h.c.\,,  \\
 \label{theta4}
{\cal{L}'''}_{KN\Theta}\, &=&\,i  
            \frac{f'_{KN\Theta}}{m_K} \overline{\Theta^+_\mu} \gamma^5
	    (p\partial^\mu K^0 + n\partial^\mu K^+)  + h.c.\,,  
\eeq
where $\Theta^+_\mu$ is the Rarita-Schwinger representation of a 
spin 3/2 state. The coupling constant $g',f,f'$ are related to the width via
\beq \label{gamma2}
    \Gamma_\Theta &=& {g'}^2_{KN\Theta} \frac{p_N (p^0_N +m_N)}
                        {2\pi m_\Theta} \,,\\
                  &=& f_{KN\Theta}^2  \frac{p^3 (p^0_N +m_N)}
		      {6\pi m_K^2 m_\Theta} \,,\\
     \label{gamma4}  &=& {f'}^2_{KN\Theta}  \frac{p^3 (p^0_N -m_N)}
		      {6\pi m_K^2 m_\Theta} 
\eeq
leading to the  values given in Table~\ref{table1}.

\begin{table} 
  \caption{Coupling strengths and cut-off parameters assuming two
    different decay widths $\Gamma_\Theta$.
    The last two columns give the positive and negative limits
    of the coupling constants for the 
    $K^*$ exchange in Eqs.~(\ref{theta1v},\ref{theta3v}) 
    which do not increase the cross section by more than a factor of two.
  }
\begin{center}
 \vspace*{10mm}
 \begin{tabular}{|c||c|c||c|c|c|c|}
 \hline
  $J^\pi$     &  \multicolumn{2}{|c||} {$\Gamma_\Theta$ = 1 MeV} & 
                \multicolumn{4}{|c|} {$\Gamma_\Theta$ = 10 MeV} \\
     &  g(f) & $\Lambda$ (GeV) & g(f) & $\Lambda (GeV)$ & 
      \multicolumn{2}{|c|}{$g^*(f^*)$} \\
  \hline
  1/2$^+$  & 1.03   &    1.03 &  3.27  &  0.76 &   0.3 & -0.6\\
  1/2$^-$  & 0.14   &    1.55 &  0.44  &  0.94 &   0.6 & -0.6\\
  3/2$^+$  & 0.46   &    0.90 &  1.47  &  0.70 &   2.0 & -0.9\\
  3/2$^-$  & 3.44   &    0.69 & 10.9   &  0.60 &   3.1 & -3.5\\
\hline
\end{tabular}
\end{center}
\label{table1}
\end{table}

Also other processes can contribute to the production. As an example we 
include the $K^*$ exchange in addition to the K exchange. 
For the positive parity
state of a spin $1/2$ baryon $B$ ($\Sigma$ or $\Theta^+$) we use
\be  \label{theta1v}
{\cal{L}}_{1/2^+}\, =\,  
            g^*_{K^*NB} \overline{B} 
	    (\gamma^\mu + \frac{\kappa}{m_\Theta+m_N} 
	    \sigma^{\nu \mu} \partial_\nu) K^{0*}_\mu \, p
		      + h.c.
\ee
and for the $3/2^+$ state
\be
 \label{theta3v}
{\cal{L}}_{3/2^+}\, =\, i 
           \frac{f^*_{K^*N\Theta}}{m_{K^*}} \overline{\Theta^{+\mu}} \gamma^5
	    \gamma^\nu p \;(\partial_\nu K^{0*}_\mu- \partial_\mu K^{0*}_\nu)  
	    + h.c.\,.  
\ee
For the negative parity states we insert the factor $i\gamma^5$ 
into Eq.~(\ref{theta1v}) and remove this factor from Eq.~(\ref{theta3v}).
The coupling constants for the $\Sigma N$ coupling  in Eq.~(\ref{theta1v})
$g^*_{K^*N\Sigma}=-3.25\sqrt{2}$ and $\kappa=1.8$ are chosen in accordance
with \cite{LiuKo,LiKo}. For the $\Theta^+$ particle 
the coupling constants will be fixed later.

\begin{figure}[t]
\begin{center}
\includegraphics[width=100mm,height=60mm]{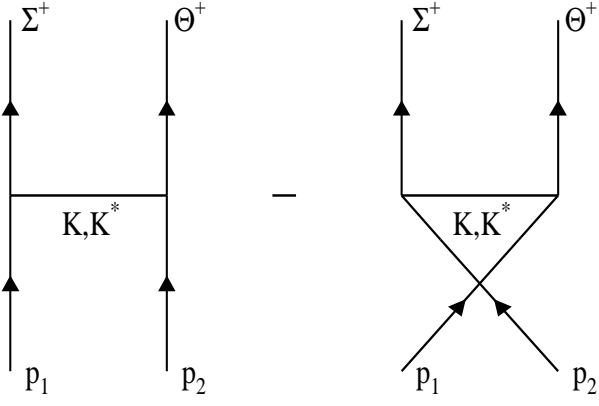}  
\end{center}
\caption{Calculated diagrams for $\Theta^+$ production in a $pp$ collision.}
\label{diagram}
\end{figure}

Accordingly to the 
diagrams shown in Fig.~\ref{diagram} we calculate the T matrix  
from which the differential cross sections is calculated
in the center-of-mass system 
\be  \label{sigma1}
\frac{d\sigma}{d\Omega}\,=\,  \frac{p_\Theta}{64 \pi^2 s p_1} \frac{1}{4}
      \sum_{\textrm{spins}}  T^*_{s_1,s_2,s_\Sigma, s_\Theta}
      (1+{\bf P}_1 {\vec{\sigma}}_1)\,
     (1+{\bf P}_2 {\vec{\sigma}}_2)\,  T_{s_1,s_2,s_\Sigma, s_\Theta}
\ee
as a function of the incoming momentum $p_1$, the outgoing momentum $p_\Theta$ and 
the center-of-mass energy $\sqrt{s}$.
The polarization of the incoming (target) proton is  described by the
vector  ${\bf P}_{1(2)}$, and ${\vec \sigma}_{1(2)}$ is the 
Pauli matrix which acts on the first (second) 
spin index, respectively.

For convenience we choose our coordinate system such that the 
z-axis coincides with the direction of the incoming
proton.  The y-axis is chosen  orthogonal to the 
reaction plane in the direction of the normal 
vector ${\bf p_1}\times {\bf p_\Theta}$
and the x-axis points to the side direction of ${\bf n}\times {\bf p_1}$.
From general considerations \cite{bystricky,Meyer} 
one can show that the differential 
cross section can only be a function of the following combinations 
of the polarization vectors
\beq  \nonumber
\frac{d\sigma}{d\Omega}\,&=&\, \big( \frac{d\sigma}{d\Omega}\big)_0
\Big( 1 + A_{y0} P_{1y}+ A_{0y} P_{2y}
  + A_{yy} P_{1y}P_{2y} + A_{xx}  P_{1x}P_{2x} \\
  &&   + A_{xz} P_{1x}P_{2z} + A_{zx} P_{1z}P_{2x} 
   + A_{zz}P_{1z}P_{2z} \Big)\,.
   \label{azimut1}
\eeq
Here the symbols $P_x,P_y,P_z$ stand for the components 
of the polarization vectors ${\bf P_{1(2)}}$.
The first factor in Eq.~(\ref{azimut1}) is 
the differential cross section for unpolarized protons. 
The coefficients $A$ depend on the polar angle $\Theta$.

Splitting the polarization vectors into components  parallel and perpendicular 
with respect to the beam direction we rewrite Eq.~(\ref{azimut1}) as
a function of the azimuth angle $\phi$ as
\beq  \nonumber
\frac{d\sigma}{d\Omega}\,=\, \big( \frac{d\sigma}{d\Omega}\big)_0&\Big(
  1 &+  A_{y0} \, P_{1\perp} \sin(\phi-\alpha_1) +
          A_{0y} \, P_{2\perp} \sin(\phi-\alpha_2) \\
     \nonumber
    & &+  A_{yy} \, P_{1\perp}P_{2\perp}
        \sin(\phi-\alpha_1)\sin(\phi-\alpha_2)  \\
   \label{phidep}
    & & + A_{xx}  \, P_{1\perp}P_{2\perp}\cos(\phi-\alpha_1)
                        \cos(\phi-\alpha_2) \\
       \nonumber
    & & + A_{zz}  \, P_{1z}P_{2z}\\			
       \nonumber
    & & + A_{xz}\,P_{1\perp}P_{2z}\cos(\phi-\alpha_1)
       + A_{zx}P_{1z} P_{2\perp}\cos(\phi-\alpha_2) \big)\;.
\eeq
Here $\alpha_1,\alpha_2$ denote  the angles of the
projections of the polarization vectors onto 
the x-y plane.
The coefficients $A_{ij}$ 
are usually called spin correlation parameters
which depend on the azimuth angle $\phi$ only up to
second order in $\cos(\phi)$ or $\sin(\phi)$.
Since the initial state is symmetric
with respect to the z-axis the parameters $A_{xx}$, $A_{yy}$, $A_{zz}$ 
are forward-backward symmetric while the others obey the 
relations 
\beq
 A_{zx}(\Theta) &=& -A_{xz}(\pi-\Theta)\, \\
 A_{0y}(\Theta) &=& -A_{y0}(\pi-\Theta)\
\eeq
similar to those for $pp\to pp\pi$ collisions \cite{Meyer}.
In our calculations using  lowest order  perturbation theory the 
coefficient $A_{y0}$ vanishes.  Thus,
we do not obtain azimuthal asymmetry if the proton 
in the beam or in the target is unpolarized.

\begin{figure}[t]
\begin{center}
\includegraphics[width=120mm, height=140mm]{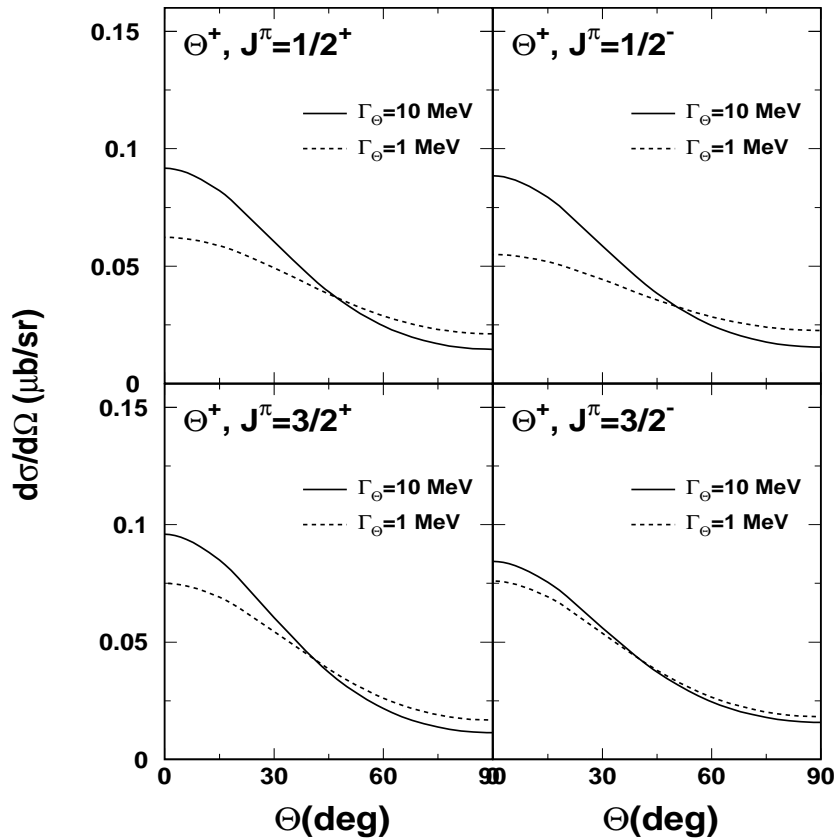}  
\end{center}
\caption{Calculated angular distributions in the center-of-mass system 
         for unpolarized beam and target for different spin-parity
	 assignments of the $\Theta^+$ states at an energy
	 of 0.13 GeV above threshold.}
\label{difsig}
\end{figure}			

\section {Threshold behavior}

The threshold behavior has widely been discussed in 
refs.~\cite{Hanhart,Nam1,thomas}. We summarize the consequences for
the correlation parameters. The produced particles can only be 
in a state with relative
orbital momentum $L=0$ since the higher partial waves are suppressed
by the centrifugal barrier. The angular distribution is isotropic 
implying $A_{xz}=0$.
Furthermore,  parity conservation
and Pauli principle imply that 
the total spin of the incoming protons is $S=0$ for positive parity 
of $\Theta^+$, and $S=1$ for negative parity. 

If we set the polarization vectors ${\bf P}_1={\bf P}_2$ and 
average over the direction of the polarization vectors we obtain from 
Eq.~(\ref{phidep}) the averaged integrated cross section
\be \label{intcross1}
\langle \sigma \rangle = 
\sigma_0 \big(1 + \frac{1}{3}P^2 (A_{xx}+A_{yy}+A_{zz})\big).
\ee
The same procedure can be applied to the polarization operator 
in Eq.~(\ref{sigma1}) leading to 
\be 
\langle \sigma \rangle  \sim 1 +  \frac{1}{3}P^2  ( {\vec{\sigma}_1}
                 {\vec{\sigma}_2} )\,.
\ee
Comparing these two last equations one arrives at the relation 
$A_{xx}+A_{yy}+A_{zz} = -3 (1)$
for $S=0(1)$, i. e. for positive (negative) parity of $\Theta^+$.

This causes the relation 
$A_{xx}=A_{yy}=A_{zz}=-1$ in the former case, 
while in the latter only the relations 
$A_{zz} = 1 - A_{yy}-A_{xx}$ and $A_{yy}=A_{xx}$
can be derived 
in a model independent way since in this case different 
spin projections contribute to the $T$ matrix. For negative parities
one obtains $A_{xx}=A_{yy}=A_{zz}=\frac{1}{3}$  
if the kaon exchange of Eqs.~(\ref{theta2},\ref{theta4}) is used.
But in general the coefficients depend on the parameters of the 
interaction  used.

\section {Results}
\subsection { $K$ exchange}

Here we consider only the $K$ exchange and calculate 
the differential cross section for the various possibilities
of the $J^\pi$ assignments.
We also investigate the effect
of the width of the $\Theta^+$ assuming values of 10 MeV and 1 MeV  which
are related to corresponding values of the coupling constants given in 
Eqs.(\ref{gamma1}, \ref{gamma2}-\ref{gamma4}). 
As a constraint we vary the cut-off parameter
$\Lambda$ such that we obtain a cross section
of  about  0.4 $\mu$b  at an excess energy of 
$\Delta E = \sqrt{s} - \sqrt{s_{\rm{thr}}}$ = 0.13 GeV 
above threshold.  This value corresponds to the result
obtained in a recent COSY measurement \cite{COSY}. 
The values fulfilling this requirement are given in Table~\ref{table1}.

\begin{figure}[t]
\begin{center}
\includegraphics[width=120mm, height=140mm]{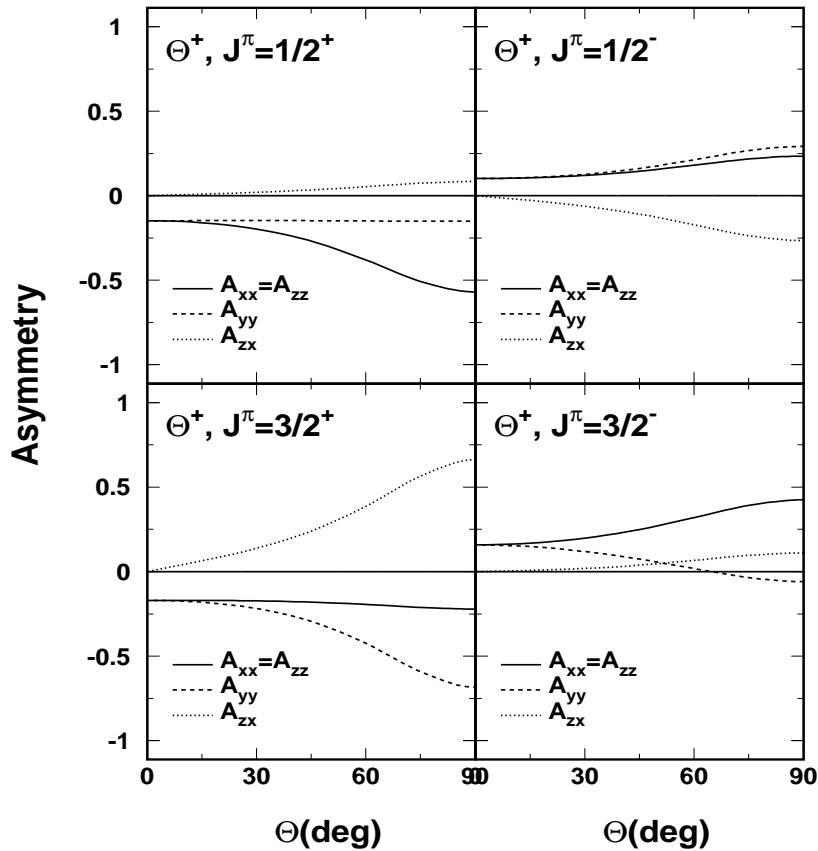}  
\end{center}
\caption{Calculated correlation parameters as a function of the polar angle
 $\Theta$ for different assumptions on the spin-parity assignments 
 of the $\Theta^+$ particle at an excess energy of 0.13 GeV.}
\label{A_theta}
\end{figure}

In Fig.~\ref{difsig} we present the angular  distribution 
$(d\sigma /d\Omega)_0$ 
for unpolarized protons in the forward region.
The cut-off parameter influences the shape of the angular
distribution. 
A strong cut-off formfactor
(small $\Lambda$ value) leads to a rather pronounced  maximum in forward
and backward direction. 

\begin{figure}[t]
\begin{center}
\includegraphics[width=120mm, height=140mm]{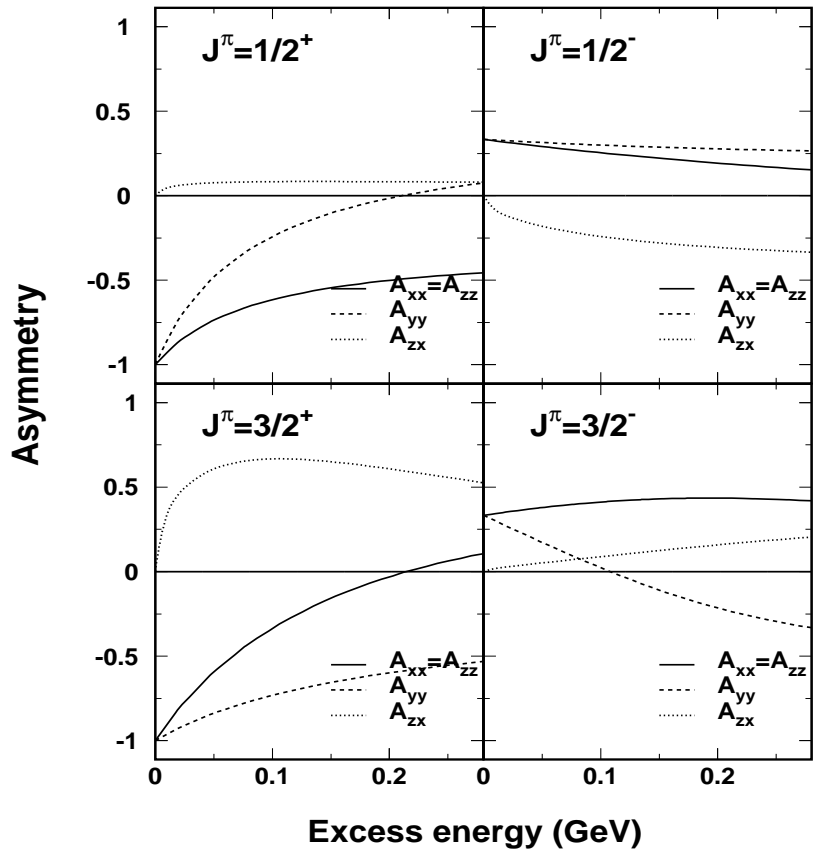}  
\end{center}
\caption{Spin correlation parameters at 90 degrees in the
 center-of-mass system
 as a function of the excess energy
 $\sqrt{s} - \sqrt{s_{\textrm{thr}}}$ for different assumptions on the
 spin-parity assignments of the $\Theta^+$ particle.}
\label{kaon}
\end{figure}

In Fig.~\ref{A_theta} we show the correlation parameters  $A_{ii}$ in
Eq.(\ref{phidep}). 
These parameters are independent of the used formfactor. 
Most of them reach their maximum value at $90^o$. 
The side-side correlation parameter $A_{xx}$ and the longitudinal
correlation parameter $A_{zz}$ coincide and $A_{zx}=-A_{xz}$
holds. 
Comparing the calculations with 
the opposite parities one recognizes that the 
transverse correlations $A_{yy}$ and $A_{xx}$ correlate their signs
in coincidence with the assumed parity of the $\Theta^+$ state. Measuring
these coefficients could give a unique signal for the determination of the 
parity as was already found in refs. \cite{Hanhart,Nam1,thomas}.
A large positive normal-long correlation parameter $A_{zx}$ and a 
negative normal-normal correlation could signalize 
the $J^\pi=3/2^+$. 

The energy dependence of the correlation parameters is shown in 
Fig.~\ref{kaon}. It is seen that the characteristics of the threshold
extend up to 50 MeV a value that has also 
been estimated in ref.~\cite{Hanhart}.
\subsection { $K^*$ exchange}

Now we investigate the effect of 
additionally including the $K^*$ exchange into the interaction. 
We treat the  coupling constants in the Lagrangians
(\ref{theta1v},\ref{theta3v}) as free parameters. To reduce 
the parameter space we choose the tensor coupling $\kappa=0$. This
is not completely unrealistic, see \cite{close2}. Furthermore we
relate the cut-off parameter to that of the $K$-exchange via
$\Lambda_{K^*}=\Lambda_K + m_{K^*} -m_K$.
Then we vary the 
values of the parameters $g^*$ and $f^*$ such that the cross section
of 0.4 $\mu$b is increased to a maximum value of  0.8 $\mu$b. 
We made calculations with both possibilities of the 
signs of the coupling constants $f^*$ or $g^*$ leading 
to constructive or destructive interference between $K$ and $K^*$
exchange. The
last two columns in Table~\ref{table1}
give the two possible values for the coupling coefficients. 
 
\begin{figure}[t]
\begin{center}
\includegraphics[width=120mm, height=140mm]{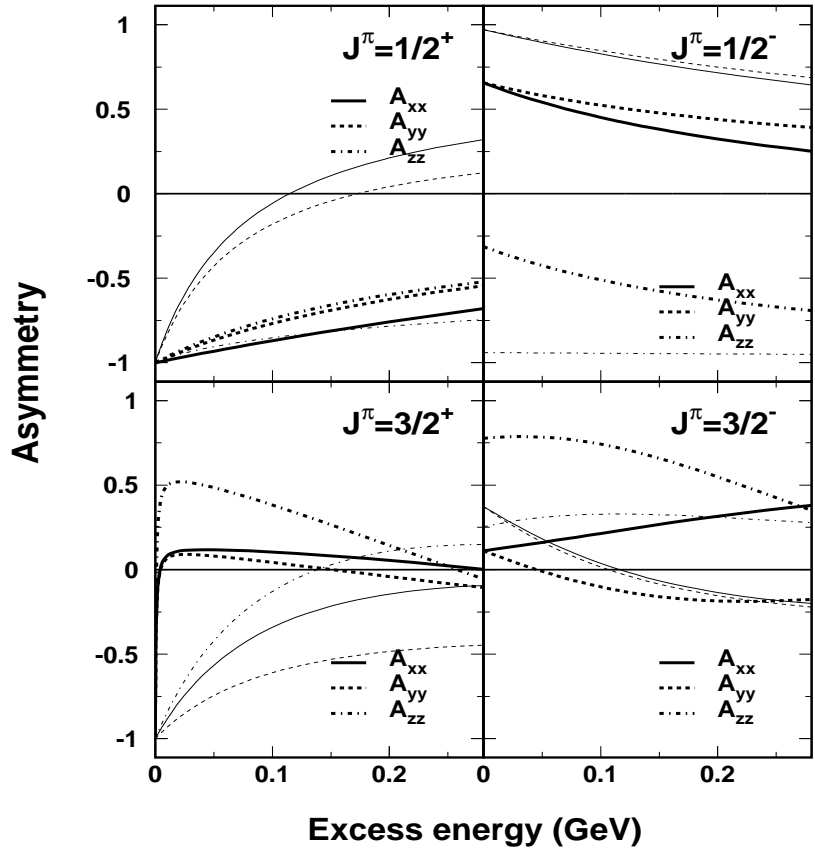}  
\end{center}
\caption{Spin correlation parameters at 90 degrees 
  as a function of the excess energy for a combination 
 of $K$ and $K^*$ exchange. The thick (thin) lines
 present calculations with
 positive (negative) coupling strengths of the $K^*$ exchange 
 given in Table~\ref{table1}.}
 \label{kstar}
\end{figure}			

In Fig.~\ref{kstar} we show the energy dependence of the
spin correlation parameters $A_{ii}$. Comparing Figs.~\ref{kaon}
and \ref{kstar} one recognizes that
the correlation parameters depend stronger on energy if the 
$K^*$ exchange has been included. The coefficients $A_{xx}$ and
$A_{zz}$ do not coincide anymore.  In particular  
 $A_{xx}$ and $A_{yy}$ of the $1/2^+$ state already change their
signs  at an excess energy of 100 MeV. A more drastic change of these
coefficients is seen for the $3/2^+$ state. 
This seems  to contradict  the estimates made in ref.~\cite{Hanhart}.
The reason for this effect 
lies in the strong destructive interference between the interaction
Lagrangians which reduces the cross section much stronger
than one expects from the $\sqrt{\Delta E}$ threshold behavior.
The constructive interference (thin lines, $f^*=-0.9$) does not show this 
strong energy dependence and agrees well with the estimates
of the behavior of the threshold region given in ref.~\cite{Hanhart}.

The behavior of the correlation coefficients for the negative states 
depends sensitively on the interaction used as can be seen by 
comparing Figs.~\ref{kaon}
and \ref{kstar}.  To identify the parity needs therefore the measurement 
of all three coefficients $A_{ii}$.


\section{Conclusion}
We have analyzed the asymmetry of the angular distribution 
for $\Theta^+$ pentaquark production 
in collisions of polarized protons. Using a variety of different
interactions it was found that the characteristic 
threshold signals survive
at energies up to 50 MeV above threshold. 
Thus such measurements are a useful tool to determine spin and 
parity of the $\Theta^+$ particle. In rare cases a more rapid 
change of the correlation parameters has been found which is
accompanied with a stronger energy dependence than the expected
$\sqrt{\Delta E}$ behavior of the cross section.

\section*{Acknowledgments}
This research was supported by German BMBF grant 06DR121,
the DAAD-MTA scientific exchange program
between Germany and Hungary, and the National Fund for Scientific
Research of Hungary, OTKA T047347.

\begin{small}

\end{small}

\end{document}